\documentclass{icrc}
\usepackage{amssymb, amsmath}
\usepackage{times}
 \usepackage{graphicx} 

\begin{document}

\title{Atmospheric muon fluxes underwater as a tool to probe
the small-$x$ gluon distribution}
\author[1]{A. Misaki}
\affil[1]{Waseda University, Ookubo 3-4-1, Shinjyuku-ku, Tokyo,
169-8555 Japan}
\author[2]{T. S. Sinegovskaya}
\author[2]{S. I. Sinegovsky}
\affil[2]{Irkutsk State University, Irkutsk, 664003 Russia}
\author[3]{N. Takahashi}
\affil[3]{Hirosaki University, Hirosaki, 036-8561 Japan}

\correspondence{sinegovsky@api.isu.runnet.ru} \runninghead{Misaki,
A. et al.: Atmospheric muon fluxes underwater} \firstpage{1}
\pubyear{2001}


\maketitle

\begin{abstract}
We compute deep-sea energy spectra and zenith-angle distributions
of the atmospheric muons, both conventional and prompt. The prompt
muon contribution to the muon flux
underwater is calculated taking into consideration predictions
of recent charm production models in which probed is the small-$x$
behavior of the gluon distribution inside of a nucleon. We argue
for a possibility to discriminate the PQCD models of the charm
production differing in the slope of the gluon distribution,
in measurements with neutrino telescopes of the
muon flux at energies 10-100 TeV.
\end{abstract}

\section{Introduction}
A correct treatment of the charm hadroproduction is
important to the atmospheric muon and neutrino
studies  since  short-lived charmed particles,
$D^\pm$, $D^0$, $\overline{D}{}^0$, $D_s^\pm$, $\Lambda_c^+$,
produced in collisions of cosmic rays with nuclei of the air,
become the dominant source of atmospheric muons and
neutrinos at energies $E \sim 100$ TeV. Thus, one needs to take
them into consideration as the background for extraterrestrial
neutrinos \citep{mann}.
 Muons originating from decay of
these charmed hadrons are so called prompt muons (PM) that
contribute to the total atmospheric muon flux.

Another aspect of the interest to the charm production
relates to the gluon density at small gluon momentum fraction $x$.
The gluon density at small $x$ is of considerable importance because
this influences strongly the charm production cross section,
both total and inclusive.  Recently \citet{prs99} and
\citet{ggv2,ggv3} have analyzed the influence of small-$x$
behavior of the parton distribution functions (PDFs) to the
atmospheric lepton fluxes at the sea level.
Basing on  next-to-leading order (NLO) calculations  of the perturbative
Quantum Chromodynamics (PQCD), they predict  PM fluxes
at the sea level depending strongly on proton gluon distributions
at small $x$ scale,  $x < 10^{-5}$.

The muon spectra underwater computed with the mo\-del of \citet{prs99},
in which used were the MRSD$_-$ \citep{MRSD-} and the CTEQ3M \citep{CTEQ3}
sets of PDFs,  were recently  discussed \citep{NSS00,note00}.
In this talk, using predictions of more recent PQCD  model \citep{ggv2,ggv3}
of the charm production,  we discuss the PM contribution to muons underwater
at depths typical for operating and constructing neutrino telescopes
\citep{Amanda,Antares,nt36,Nestor}.
Namely, we try to study a PM flux underwater dependence
on the slope $\lambda$ of the gluon PDF at small $x$:
 $xg(x)\propto x^{-\lambda}$.
The nature of the small-$x$ behavior of the gluon density is now under
extensive discussion (see, for example,
 \citet{ball00,schl,vogt00,yosh01}).
The  small-$x$ behavior of the PDFs is the subject of the deep interest
since the realization of the underlying dynamics is yet far from
being complete.

\section{PDFs and charm production models}
Due to dominant subprocess in heavy quarks hadroproduction,
$gg\rightarrow c\overline c$, the charm production is sensitive to
the gluon density at small $x$, where $x$ is the gluon momentum fraction.
One may evaluate the scale of $x$ in cosmic ray
interactions as follows.  The product of the gluon momentum fraction $x_{1}$
of the projectile nucleon and that of the target $x$ near
the charm production threshold ($\sim 2m_{c}$) is \,
\mbox{$x_{1}x=4m_{c}^{2}/(2m_{N}E_{0})$},\,\
 where $E_{0}$ is the primary
nucleon energy in the lab frame. Since a muon takes away about 5\% of the
primary nucleon energy, $E_{0}\simeq 20E_{\mu}$, we have
\mbox{$x_{1}x=0.1(m_{c}/m_{N})(m_{c}/E_{\mu})$.}
Because of the steepness of the primary cosmic ray spectrum
only large $x_{1}$ contribute sizeably to the atmospheric charm
production, one needs adopt $x_{1}\gtrsim 0.1$.
Taking $m_{c}^{2}\simeq 2\,$ GeV$^{2}$, one may find for
$E_{\mu}\gtrsim 100$\,TeV  the range of importance to be
$x\lesssim 2\cdot10^{-6}$.
It should be stressed, this range is yet  outside of the scope
of the perturbative next-to-leading order global analysis
of parton distributions \citep{MRST,CTEQ5}.

The spectral index $\lambda$ being used in the PQCD charm production
models~\citep{prs99,ggv3} covers wide range from about
$0.1$, the number that may be connected to the soft pomeron
with intercept $1+\epsilon_{0}$ (where $\epsilon_{0}=0.08$),
to about $0.5$ in the 3-flavor scheme of the  BFKL approach
\citep{BFKL}.
 \begin{figure}[t]
  \vskip -10mm
 \vspace*{2.0mm} 
 \includegraphics[width=8.3cm]{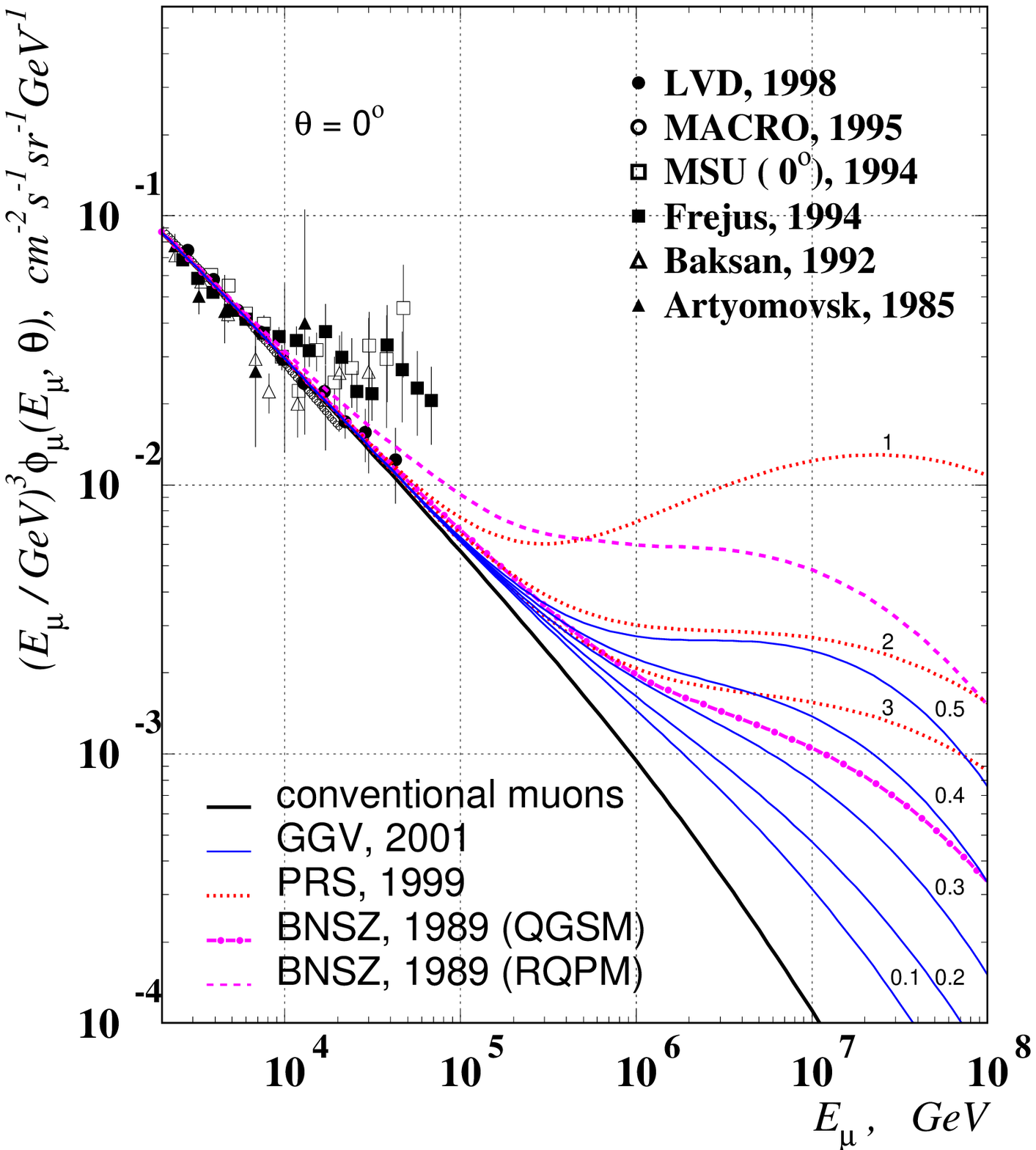} 
\caption{Vertical sea-level muon flux predictions.
 (See~\citet{prd58,note00} for data references.)
 \label{fig-1}}
  \vskip -3mm
 \end{figure}

In Fig.~\ref{fig-1} shown are sea-level PM fluxes (added to the
conventional muon flux) predicted with PQCD charm production models
by \citet{prs99} (hereafter PRS) and by \citet{ggv2,ggv3} (GGV),
which are used in our calculation of the deep-sea muon flux.
Let us sketch out these models.

\begin{figure}[t!]
 \vskip -8mm
 \vspace*{2.0mm} 
 \includegraphics[width=8.3cm]{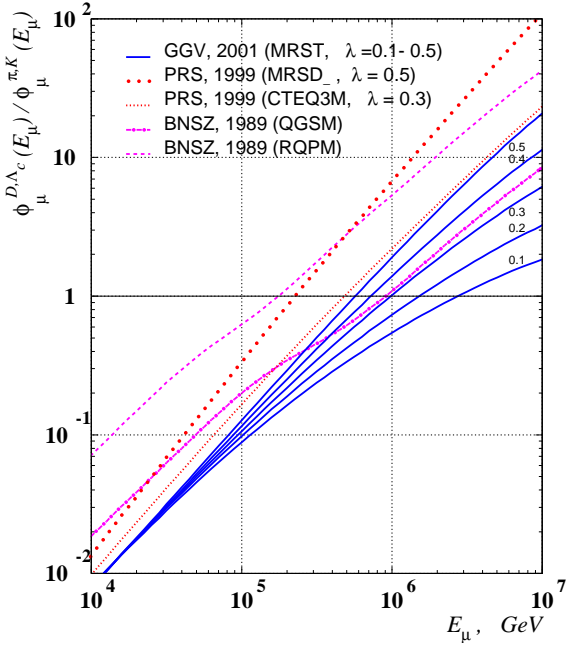} 
\vskip -2mm
 \caption{Sea-level ratio of the differential prompt muon flux to the
 conventional one.
 \label{fig-2}}
 \end{figure}

 \underline{PRS-1}:\\[0.1cm]
The PRS-1 model (dot lines in Figs.~\ref{fig-1}, \ref{fig-2})
(identical with the PQCD-1 in Ref.~\citep{note00}) is
based on the  MRSD$_-$ set \citep{MRSD-}.
The PDF input parameters are the followings:
$xg(x,Q_0^2)\sim x^{-0.5}$ as $x\rightarrow0$,
4-momentum transfer squared $Q_0^2=4$\,GeV$^2$;
the sea light quark asymmetry, $\overline u < \overline d$,
is taking into consideration; the QCD scale in the minimal subtraction scheme
($\overline {\rm MS}$), $\Lambda^{\overline {\rm MS}}_4=0.215$\,GeV,
corresponds to the effective coupling at the $Z$ boson mass scale
$\alpha_{s}(M^{2}_{Z})=0.111$.  The factorization scale is
$\mu_F=2m_c$, the renormalization one is $\mu_R=m_c$, where the
charm quark mass, $m_c$, is chosen to be equal $1.3$.
The sea-level prompt muon flux was parameterized by authors \citep{prs99} as
\begin{eqnarray}\label{prs1}
\log_{10}[E^3_{\mu}\phi^{D,\Lambda_{c}}_{\mu}(E_{\mu})/
(\mathrm{cm^{-2}s^{-1}sr^{-1}GeV^{2}})] \nonumber \\
\ \ \ \ \ \ \ \ =-5.91+0.290y+0.143y^{2}-0.0147y^{3},
 \end{eqnarray}
\noindent
 where $y=\log_{10}(E_{\mu}/{\rm GeV}).$

 \underline{PRS-2}:\\[0.1cm]
In the PRS-2 model (identical with the PQCD-2 in  Ref.
\citep{note00}), CTEQ3M set~\citep{CTEQ3} was used.
Corresponding inputs that were utilized in this model are
$\Lambda^{\overline{MS}}_4=0.239$\, GeV, 
$\alpha_{s}(M^{2}_{Z})=0.112,$ \,
 $m_c=$ 1.3 GeV, \,
$\mu_F=2m_c$,  $\mu_R=m_c$,\, and  $\lambda=0.286$\,
at $Q^{2}_{0}=1.6\,{\rm GeV}.$
The corresponding PM flux was parameterized as
\begin{eqnarray}\label{prs2}
\log_{10}[E^{3}_{\mu}\phi^{D,\Lambda_{c}}_{\mu}(E_{\mu})/
(\mathrm{cm^{-2}s^{-1}sr^{-1}GeV^{2}})] \nonumber \\
\ \ \ \ \ \ \ \  =-5.79+0.345y+0.105y^{2}-0.0127y^{3}.
\end{eqnarray}

 \underline{PRS-3}:\\[0.1cm]
In this model the CTEQ3M set was also used.
Differing  from PRS-2 in renormalization and factorization
scales, $\mu_F=\mu_R=m_c$, this model appears the uncertainty
due to the scale choice. The PM spectrum predicted is now
\begin{eqnarray}\label{prs3}
\log_{10}[E^3_{\mu}\phi^{D,\Lambda_{c}}_{\mu}(E_{\mu})/
(\mathrm{cm^{-2}s^{-1}sr^{-1}GeV^{2}})] \nonumber \\
\ \ \ \ \ \ \ \ =-5.37+0.0191y+0.156y^{2}-0.0153y^{3}.
\end{eqnarray}

\underline{GGV}:\\[0.1cm]
Here we present results for the model, among those discussed
by \citet{ggv3}, which is based on MRST set of  PDFs
\citep{MRST} with different values of the slope $\lambda$ in the range
$0.1-0.5$, \,$Q^{2}\geq 1.25$\,GeV$^{2}$;\, $\alpha_{s}(M^{2}_{Z})=0.1175.$
The factorization and renormalization scales are:
$$
  \mu_F=2m_T, \,\mu_R=m_T, \,
$$
where
$$
m_{T}=(k_{T}^{2}+m_c^{2})^{1/2},\,
  m_c= 1.25\, {\rm GeV},
$$
and characteristic transverse momentum $k_{T}$
is of $\sim m_c.$

In order to compute PM flux underwater we
parameterize with Eq.~(\ref{pms}) sea-level muon spectra
of the GGV model (see Fig. 7 in  Ref.~\citep{ggv3}):
\begin{eqnarray}\label{pms}
\phi_{\mu}^{D,\Lambda_{c}}(E_\mu)/(\mathrm{cm^{-2}s^{-1}sr^{-1}GeV^{-1}})
\nonumber \\
\ \ \ \ \ \ \ \ = A\left(E_\mu/{\rm GeV}\right)^{-(\gamma_{0}+\gamma_{1}y+
\gamma_{2}y^{2}+\gamma_{3}y^{3})} \,.
\end{eqnarray}
\begin{table}[h!]
\vskip -0.5 mm
\protect\caption{$\lambda$ dependent parameters of the prompt muon
 differential spectra at the sea level (Eq.~(\ref{pms})).
\label{t1}}
\center{\begin{tabular}{cccccc} \hline\hline
$\lambda$ &$A, 10^{-6}$&$\gamma_{0}$&$\gamma_{1}$&$\gamma_{2},
10^{-2}$&$\gamma_{3}, 10^{-3}$\\ \hline
$0.1$    &$ 3.12$ &$2.70 $&$-0.095$ &$1.49$&$-0.2148$ \\
$0.2$    &$ 3.54$ &$2.71 $&$-0.082$ &$1.12$&$-0.0285$ \\
 $0.3$    &$ 1.80$ &$2.38 $&$0.045$&$- 0.82$ &$0.911$ \\
 $0.4$    &$ 0.97$ &$2.09 $&$0.160$&$- 2.57$ &$1.749$
\\ $0.5$    &$ 0.58$ &$1.84 $&$0.257$&$- 4.05$ &$2.455$ \\
\hline\hline \end{tabular}}
 \end{table}
In Table~\ref{t1} five sets of the parameters to Eq.~(\ref{pms}) are
presented for different values of the small-$x$ gluon PDF
spectral index $\lambda$.

\begin{figure}[t]
 \vskip -17mm
 \vspace*{2.0mm} 
 \includegraphics[width=8.3cm]{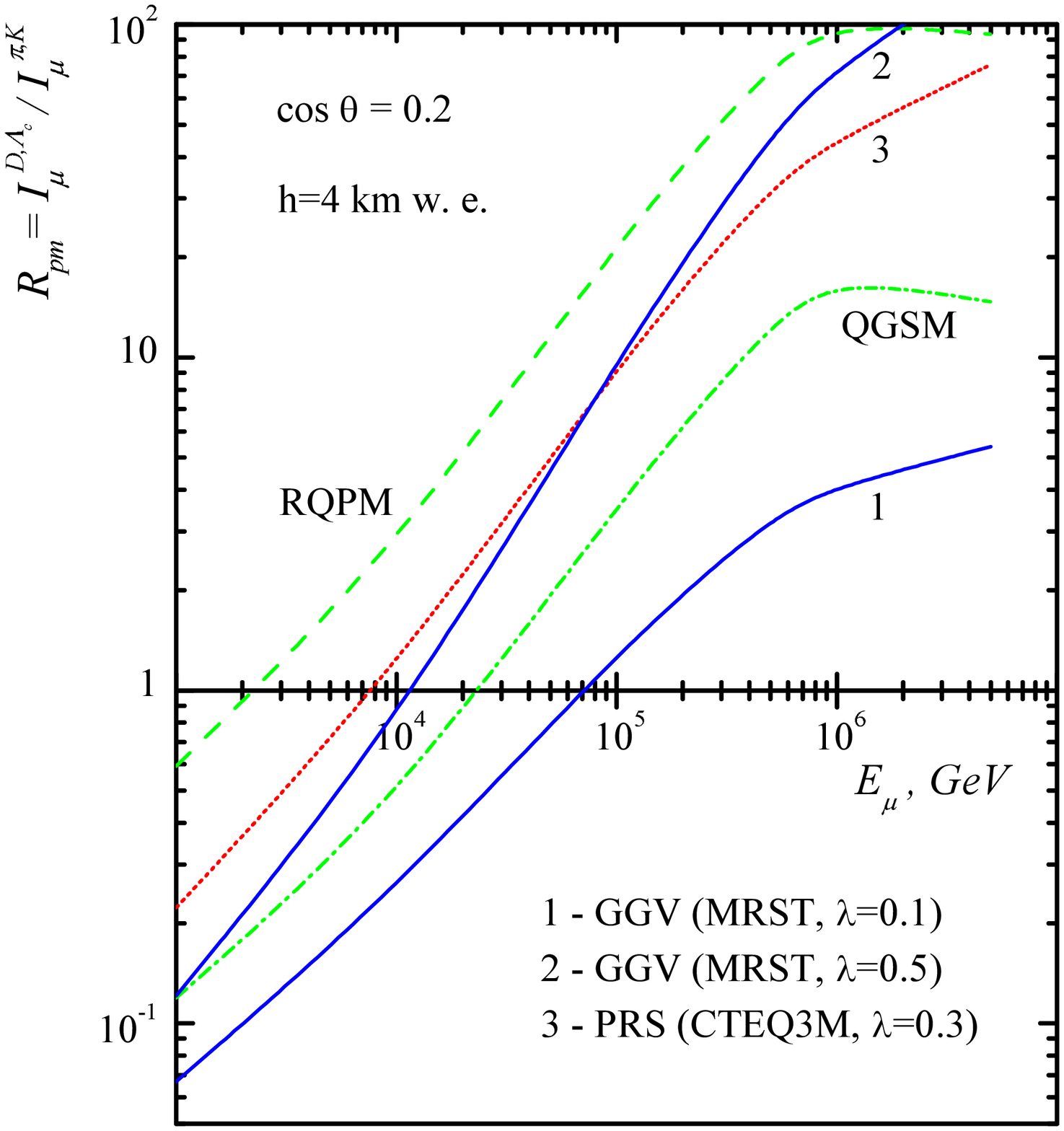} 
 \vskip -19mm
 \caption{Prompt muon contribution at $h=4$\,km w. e. vs. $E_{\mu}$.
   \label{fig-4}}
 \vskip -0 mm
    \end{figure}

We calculated the conventional muon flux basing on the  nuclear cascade
model by \citet{VNS86} (see also \citet {NSS98,prd58}).
High-energy part of this spectrum for the vertical may be approximated
with formula (in $\mathrm{cm^{-2}s^{-1}sr^{-1}GeV^{-1}}$)
\begin{equation}\label{our}
\phi_\mu^{\pi,K}\left(E_\mu,0^\circ \right) =
  \begin{cases}
 14.35\,E_\mu^{-3.672} &
 \text{$E_1<E_\mu\leqslant E_2$}, \\
 10^{3}E_\mu^{-4}& \text{$E_\mu>E_2$},
  \end{cases}
\end{equation}
where $E_1=1.5878\times10^3\,\mathrm{GeV}, E_2=4.1625\times10^5$\,GeV.

Five thin lines in Fig.~\ref{fig-1} present the sum of the conventional
muon flux Eq.~(\ref{our}) and the GGV PM flux (Eq.~(\ref{pms})) corresponding
to $\lambda=0.1, 0.2, 0.3, 0.4, 0.5$ (numbers near lines). Dot lines
show the same for  PRS models, Eqs.~(\ref{prs1}-\ref{prs3}).
For comparison there are also shown
contributions due to the quark-gluon string
model (QGSM) and the recombination quark-parton one (RQPM)
\citep{BNSZ89,prd58} (dash-dot and dash, respectively).
Ratios of PM fluxes according above models  to the conventional
flux are shown in Fig.~\ref{fig-2}.
As one can see, the cross energy of the PM fluxes and
conventional one covers the wide region from $\sim ~ 150$\,TeV to
$\sim ~ 3 $\,PeV, that is more than one order of the magnitude.

\section{Prompt muon component of the flux  underwater}
Muon energy spectra and angle distributions of the flux
underwater was computed with the method by~\citet{NSB94}.
The collision integral in the kinetic equation
includes  the energy loss of muons due to
brems\-strah\-lung, direct $e^+e^-$ pair production and photonuclear
interactions. The ionization energy loss and the small-$v$ part of the
loss due to $e^+e^-$ pair production ($v < 2\cdot10^{-4}$, where
$v$ is the fraction of the energy lost by the  muon)  were treated as
continuous ones.
In our calculations of underwater muon fluxes at different zenith angles,
we used PQCD PM fluxes calculated only for the vertical direction,
supposing the isotropic approximation for prompt muons to be a reliable
at least for $10^4< E_\mu < 10^6\,$ GeV  at zenith angles
$\theta\lesssim 80^\circ.$

The prompt muon fraction of the flux underwater, $R_{pm}$, defined as
ratio of the integral prompt muon spectrum to the conventional one, is
presented in Fig.~\ref{fig-4} for  the depth of 4 km of the  water
equivalent (w. e.) and for $\cos\theta=0.2$.
As is seen from the figure, $R_{pm}$ related to the
gluon density slope $\lambda=0.5$ is a factor 3 greater than
that for $\lambda=0.1$ at $E_\mu \gtrsim 10$ TeV.

\begin{figure}[t]
  \vskip -15mm
 \vspace*{2.0mm} 
  \includegraphics[width=8.3cm]{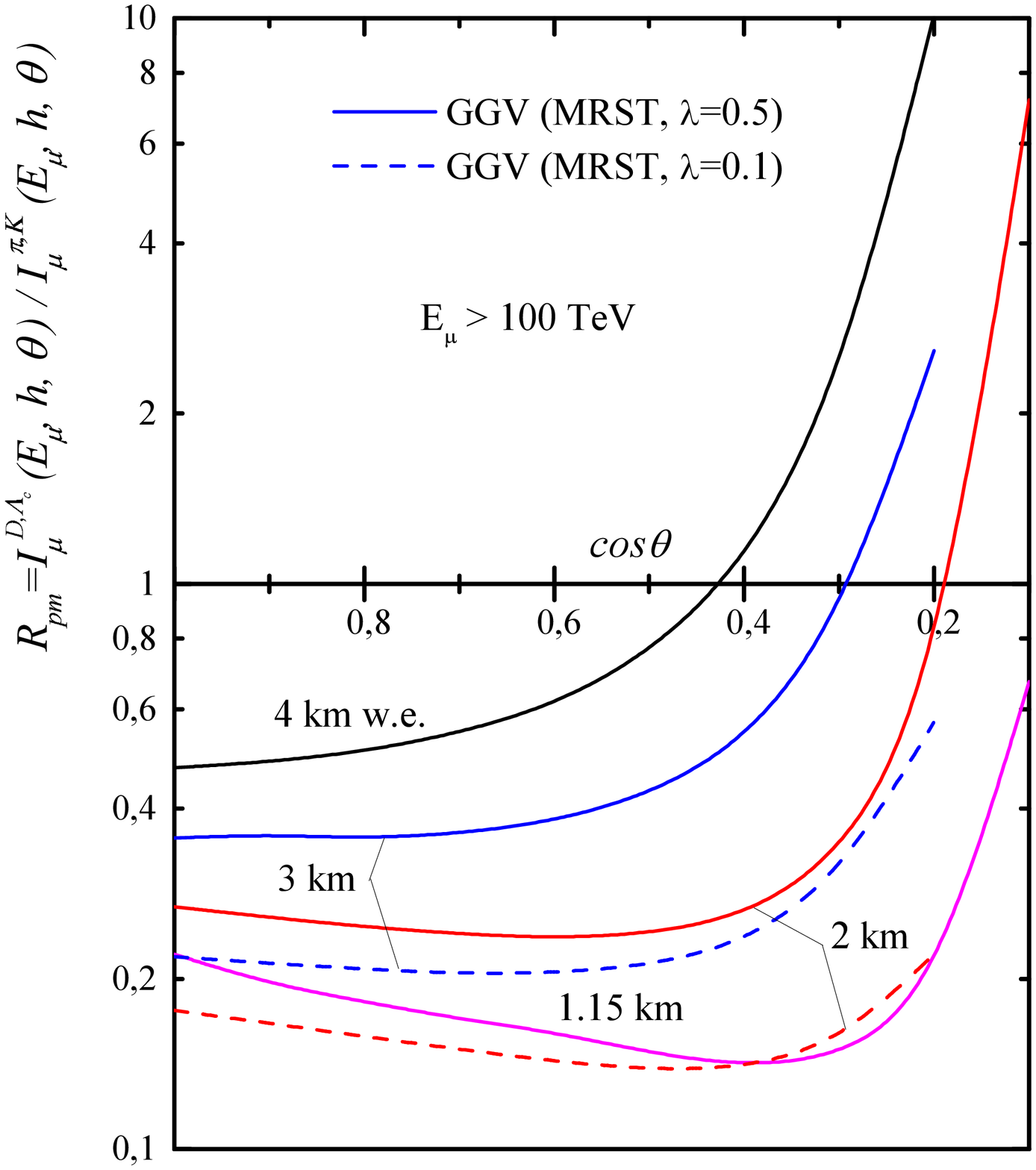} 
 \vskip -15mm
 \caption{Ratio of the prompt muon flux underwater
   to the conventional one as a function of $\cos\theta$.
   \label{fig-5}}
   \end{figure}

Zenith-angle distributions of the prompt muon contribution at
depths $1-4$\,km w. e., calculated for $E_\mu > 100$\,TeV,  are shown
in Fig.~\ref{fig-5}. Only GGV model predictions were used here
with $\lambda=0.1$ (dash) and $\lambda=0.5$ (solid).
For the vertical one can see $R_{pm}$ rising from about 0.2
at the depth of Baikal NT (1.15 km) to
about 0.5 at the NESTOR depth ($\sim 4$ km). For the larger zenith angles,
$\theta\sim 75^{\circ}$, this contribution becomes apparently sizable
at depths $3-4$\,km.  Differences in the predictions
owing to a change of  $\lambda$, from  $0.1$ to $0.5$ (see $h=2$ and
$3$ km w. e.), are also clearly visible: the ratio
$R_{pm}(\lambda=0.5)/R_{pm}(\lambda=0.1)$ at $h=2$\,km w. e. grows from
about $1.5$ to about $5$ as $\cos\theta$ changes from $1$ to $0.2$.

Here we supposed no differences between the PRS and GGV models
apart from those related to the charm production cross sections.
Actually one needs  to compare the primary spectrum and composition,
nucleon and meson production cross sections and other details of the
atmospheric nuclear cascade being used in above computations.
These sources of uncertainties would be considered elsewhere.

\section{Summary}
In order to test the small-$x$ gluon distribution effect
we have computed deep-sea prompt muon fluxes using predictions of charm
production models based on NLO calculations of the PQCD \citep{prs99,ggv3}.
The possibility to discriminate the PQCD models, differing
in the slope of the gluon distribution, seems to be achievable
in measurements of the underwater muon flux at energies 10-100 TeV.

Being hardly appeared at the sea level for energies up to $10^5$ GeV
(Figs.~\ref{fig-1}, \ref{fig-2}), a dependence on the spectral
index $\lambda$ of the small-$x$ gluon distribution becomes more
distinct at depths $3-4$\,km w. e. (Figs.~\ref{fig-4}, \ref{fig-5}).
At the depth of $4$\, km and at the angle of $\sim 78^{\circ}$ one could
observe the PM flux to be equal, for $\lambda=0.5$, to the conventional
one even for muon energy $\sim 10$\,TeV (the cross energy).
While for $\lambda=0.1$ the cross energy is about $70$\,TeV.
For the high energy threshold, $E_{\mu}>100$\,TeV, and at $h\lesssim 3$\,
km w. e., the ratio $R_{pm}$ is nearly isotropic up to $\sim 60^{\circ}$.
The ``cross zenith angle'' at a given depth, $\theta_c(h)$, depends
apparently on $\lambda$:

$\cos\theta_{c}\mid_{\lambda=0.5}\simeq0.3,\,\,\cos\theta_{c}\mid_{\lambda=0.1}
\simeq 0.1,\,\,h=3\,\, \mathrm{km\, w.\, e.}$

\begin{acknowledgements}
The work of T. S. and S. S. is supported in part by the Ministry of
Education of the Russian Federation under  Grant No.~015.02.01.004
(the Program ``Universities of Russia -- Basic Research'').

\end{acknowledgements}

\end{document}